# Giant superlinear power dependence of photocurrent based on layered Ta$_2$NiS$_5$ photodetector

Xianghao Meng[1,2,†], Yuhan Du[1,2,†], Wenbin Wu[1,2,†], Nesta Benno Joseph[3], Xing Deng[4], Jinjin Wang[4], Jianwen Ma[5], Zeping Shi[1,2], Binglin Liu[1,2], Yuanji Ma[1,2], Fangyu Yue[1,2,4], Ni Zhong[1,2,4], Ping-Hua Xiang[1,2,4], Cheng Zhang[5,6], Chun-Gang Duan[1,2,4], Awadhesh Narayan[3], Zhenrong Sun[1,2], Junhao Chu[2,4,7], Xiang Yuan[1,2*]

[1]State Key Laboratory of Precision Spectroscopy, East China Normal University, Shanghai 200241, China

[2]School of Physics and Electronic Science, East China Normal University, Shanghai 200241, China

[3]Solid State and Structural Chemistry Unit, Indian Institute of Science, Bangalore 560012, India

[4]Key Laboratory of Polar Materials and Devices, Ministry of Education, East China Normal University, Shanghai 200241, China

[5]State Key Laboratory of Surface Physics and Institute for Nanoelectronic Devices and Quantum Computing, Fudan University, Shanghai 200433, China

[6]Zhangjiang Fudan International Innovation Center, Fudan University, Shanghai 201210, China

[7]Institute of Optoelectronics, Fudan University, Shanghai 200438, China

*Correspondence and requests for materials should be addressed to X. Y. (E-mail: xyuan@lps.ecnu.edu.cn)

†These authors contributed equally to this work.

# Giant superlinear power dependence of photocurrent based on layered $Ta_2NiS_5$ photodetector

**Abstract**

Photodetector based on two-dimensional (2D) materials is an ongoing quest in optoelectronics. These 2D photodetectors are generally efficient at low illuminating power but suffer severe recombination processes at high power, which results in the sublinear power dependence of photoresponse and lower optoelectronic efficiency. The desirable superlinear photocurrent is mostly achieved by sophisticated 2D heterostructures or device arrays, while 2D materials rarely show intrinsic superlinear photoresponse. Here, we report the giant superlinear power dependence of photocurrent based on multi-layer $Ta_2NiS_5$. While the fabricated photodetector exhibits good sensitivity ( 3.1 mS/W per □ ) and fast photoresponse (31 μs) , the bias-, polarization-, and spatial-resolved measurements point to an intrinsic photoconductive mechanism. By increasing the incident power density from 1.5 μW/μm² to 200 μW/μm² , the photocurrent power dependence varies from sublinear to superlinear. At higher illuminating conditions, a prominent superlinearity is observed with a giant power exponent of $\gamma = 1.5$. The unusual photoresponse can be explained by a two-recombination-center model where the distinct density of states of the recombination centers effectively closes all recombination channels. The fabricated photodetector is integrated into camera for taking photos with enhanced contrast due to the superlinearity. Our work provides an effective route to enable higher optoelectronic efficiency at extreme conditions.

**Introduction**

Optoelectronic devices based on two-dimensional (2D) materials have attracted intense research attention owing to their excellent performances of high sensitivity[1,2], fast response time[3,4], and high electron mobility[5,6]. The photoconductive detector is one of the most stable optoelectronic devices with broad working bandwidth[7], high responsivity[8], and high gain[9]. The photoresponse of this device is mainly determined by material properties due to the simple structure and physical mechanism. When semiconductor material absorbs incident photons, whose energy is equal to or greater than the bandgap, photon-generated electrons and holes will be separated in opposite directions and collected by the electrodes with an external bias. The photocurrent ($I_{ph}$) increases as a function of incident power ($P$) following a power-law dependence of $I_{ph} \propto P^{\gamma}$. The power exponent ($\gamma$) varies between different materials because of electron-hole generation, trapping, recombination process and other mechanisms[10–13].

For the ideal case, a linear increase of photocurrent with incident power is expected ($\gamma = 1$) since the photocurrent is solely determined by the photogeneration of electron-hole pairs[14–19]. In most 2D-based devices, $I_{ph}$ exhibits a sublinear power dependence under high-intensity illumination due to dominating contribution from defects and impurities. As light intensity increases, those defects serve as effective recombination centers and capture more photocarriers which lead to the saturation of photocurrent ($\gamma < 1$)[15,20–22] and decreased responsivity. As for superlinear power dependence ($\gamma > 1$), it is found in comparatively rare cases and features increased photoresponsivity with power[23,24]. Recently, the superlinear power-dependent photocurrent was reported in a series of artificial 2D structures such as graphene/h-BN[25], graphene/WSe$_2$[26], WS$_2$/MoS$_2$[27] heterojunctions, and sheet array[28]. The typical origin of superlinearity from heterostructure devices is the photothermionic effect, where hot carriers are injected from the gate side to overcome the Schottky barrier exponentially as external injection bias increases, resulting in significantly extended spectral bandwidth and responsivity[25,26,29]. Meanwhile, the multi-center Shockley-Read-Hall process[28,30] also contributes to the superlinear response in arrayed structures such as printed MoS$_2$ and GaTe transistor arrays because the array structure keeps photocarriers from massive recombination at high luminous power[30].

The desired superlinear photoresponse is mainly achieved by sophisticated 2D heterostructures and arrays[25–29]. However, as the building block of those 2D artificial structures, the 2D materials rarely show intrinsic superlinear photoresponse. Even within the existing cases, the superlinearity is weak with power-law exponent γ generally lower than 1.1. Hereafter, we define "homogeneous 2D material"[31] as those single 2D material that contrasts the heterostructures and arrays. Homogeneous 2D material with stronger intrinsic superlinearity (higher γ) is desired which potentially allows for stronger optoelectronic efficiency at the high power regime and enables better performance if integrated into the discussed sophisticated structures.

In this work, we report the prominent and intrinsic superlinear power dependence of photocurrent based on homogeneous $Ta_2NiS_5$ at ambient condition. The photodetector manifests itself with a simple metal-$Ta_2NiS_5$-metal structure. Bias-dependent and spatial scanning photocurrent measurements suggest the photoconductive origin of the photoresponse so that the photocurrent is determined by the intrinsic material property of $Ta_2NiS_5$. The photoconductive devices feature a fast response of 31 μs, along with good sensitivity of 3.1 mS/W per □, and polarization-sensitive anisotropy. At the low intensity regime $(1.5 - 15\ \mu W/\mu m^2)$, photocurrent shows conventional sublinear power dependence. Upon increasing the power density $(15 - 200\ \mu W/\mu m^2)$, the photocurrent becomes weakly superlinear. With illuminating power density higher than $200\ \mu W/\mu m^2$, strong superlinear power dependence is found with a giant power exponent $\gamma = 1.5$ for the homogeneous 2D material. Different from the previous report[32] where the capture cross-section plays a major role in determining the weak superlinearity, here, the unusual strong superlinearity requires the presence of recombination centers with distinct density of states. We present a two-recombination-center (RC) model to capture the main finding of the experiments which is further quantitatively proved by the multi-parameter fitting. The fabricated $Ta_2NiS_5$ device is tested for taking photographs. The image contrast is clearly enhanced due to the superlinearity of the device. Our work sheds light on the superlinear photocurrent which allows enhanced optoelectronic performance of photoconductive devices at high illuminating power.

$Ta_2NiS_5$ crystallizes in the orthorhombic system (space group Cmcm, $D_{2h}^{17}$) as shown

in Fig. 1a, which is composed of layers stacking along b-axis. Each layer consists of the periodically arranged [TaS$_6$]$_2$ chains and NiS$_4$ chains. The armchair structure runs along the a-axis leading to the quasi-one-dimensional structure[33,34] along with the resultant anisotropic electronic and optical characteristics[35,36]. The high-quality Ta$_2$NiS$_5$ crystals are prepared by chemical vapor transport method (Fig. 1b) with temperature gradient of 6 °C/cm. The needle-like crystals (Fig. 1c) are found in the cold end with shiny surfaces. More details can be found in the *Method* section. As shown in Fig. 1d, the copper target X-ray diffraction (XRD) pattern of the as-grown Ta$_2$NiS$_5$ crystal is performed to evaluate the crystal structure and orientation. The prominent peaks at 14.6°, 29.4°, and 44.8° originate from the (010) plane. The extracted lattice constant $b$ is 12.11 Å. The inset presents the full width at half-maximum (FWHM) of 0.16°. The lattice properties and anisotropic characteristics can be further examined by Raman microscope. The randomly polarized Raman spectrum is shown in Fig. 1e which is measured under ambient condition with HeNe laser. Apparent peaks at 127.0 cm$^{-1}$ and 148.6 cm$^{-1}$ correspond to the $^2$A$_g$ and $^3$A$_g$ phonon modes, respectively[37]. Angle-resolved polarized Raman spectra are carried out in both parallel and perpendicular polarization configurations. Figure. 1f and Figure. 1g present the false-color maps of the Raman spectra. The original spectra are provided in Supplementary Section II. The experimental coordinate $x, y, z$ coincides with the crystal direction $a, b, c$, respectively. The excitation beam propagates in $y$ direction and the polarization is controlled by a half-wave plate. More details are provided in the *Method* section and Supplementary Section II. The Raman tensor of A$_g$ modes in Ta$_2$NiS$_5$ is given by[38]

$$R(\mathrm{A_g}) = \begin{pmatrix} |a|e^{i\varphi_a} & & \\ & |b|e^{i\varphi_b} & \\ & & |c|e^{i\varphi_c} \end{pmatrix} \quad (1)$$

The anisotropic Raman response can be quantitatively derived as

$$I_\parallel(\mathrm{A_g}) \propto |c|^2 \{(\sin^2\theta + \frac{|a|}{|c|}\cos\varphi_{ca}\cos^2\theta)^2 + (\frac{|a|}{|c|}\sin\varphi_{ca}\cos^2\theta)^2\} \quad (2)$$

$$I_\perp(\mathrm{A_g}) \propto \frac{1}{4}(|a|^2 + |c|^2 - 2|ac|\cos\varphi_{ca})\sin^2 2\theta \quad (3)$$

$a, b,$ and $c$ are the amplitude of Raman tensor elements. The $\varphi_a$, $\varphi_b$, and $\varphi_c$ are the phases of the elements, and $\varphi_{ca} = \varphi_c - \varphi_a$. $\theta$ denotes the angle between the polarization vector of incident light $e_i$ and the a-axis of the crystal[39]. The angle-dependent phonon intensity can be well fitted by the Raman tensor as shown in Fig. 1h-

k. In the parallel configuration ($e_i \parallel e_s$), $I_\parallel(A_g)$ reaches the global maximum along the armchair direction and local maximum along the zigzag direction. Meanwhile, both $A_g$ modes present four-fold symmetry in the perpendicular configuration ($e_i \perp e_s$). The polarized Raman spectra agree with the theoretical prediction and help to identify the crystal direction. Based on our infrared spectroscopy measurement, a direct band gap of 273 meV is extracted for the as grown $Ta_2NiS_5$ which agrees with the general consensus of $Ta_2NiS_5$ being a narrow gap semiconductor[36,37,40]. More details are given in Supplementary Section VI.

To examine optoelectronic properties of multi-layer $Ta_2NiS_5$, the as-grown single crystals are exfoliated by mechanical method, and device fabrications are performed by a home-built lithography system with lift-off procedures. Fig. 2a exhibits the schematic diagram of the device structure. The multi-layer $Ta_2NiS_5$ is transferred to the $SiO_2$/Si substrate and contacted by electrodes (5 nm Cr/70 nm Au). The photocurrent is measured under the ambient condition with illumination of 632.8 nm laser. Due to the narrow gap nature of $Ta_2NiS_5$[36], the photoresponse is expected to be insensitive to the wavelength of visible lasers, but the laser beam with lower wavelength is found capable of damaging the sample at moderate intensity. Fig. 2b depicts the bias-dependent photocurrent with incident power density $p = 0.324$ mW/μm² (defined as incident power per unit area). The edge of the spot size is defined by the position with 1.5 standard deviations. The photocurrent $I_{\text{ph}}$ is defined as $I_{\text{ph}} \equiv I_{\text{illumination}} - I_{\text{dark}}$, which describes the difference between current with and without laser illumination. The measured photocurrent presents symmetric and linear bias dependence and goes through the origin of the plot. The photocurrent is extracted as $I_{\text{ph}} = 5.35$ μA under bias voltage of $U = 1$ V and incident power density of $p = 0.324$ mW/μm². The photoresponsivity reaches a reasonable value of $R_\lambda = \frac{I_{\text{ph}}}{P} = 2.5$ mA/W in small-gap semiconductor[41]. The $R_\lambda$ does not reflect the intrinsic property of the device and material since it varies with the bias. A more proper physical parameter is the photoconductive responsivity $R_\sigma$ extracted as 3.1 mS/W per □. Fig. 2c exhibits the dark current which is also symmetric and linear with bias, proving the Ohmic contact of the device as an important prerequisite for high-performance devices[3]. The observed bias dependence suggests the photoconductive origin rather than the photovoltaic mechanism of the measured device. Otherwise, the Schottky barrier or other built-in potential results in the nonlinear response in both $I_{dark} - U$ and $I_{ph} - U$ test[24,42,43]. Meanwhile, the negligible photocurrent at $U = 0$ V is also against the photovoltaic mechanism. Fig. 2d is the image of the device, and the scale bar is 10 μm. The height profile (inset) is measured along the white dashed line, suggesting the thickness of 178 nm of Ta$_2$NiS$_5$ flake. The photoconductive origin is further proved by spatial-resolved experiments as shown in Fig. 2e. The photocurrent is measured along the red dash line with $U = 1$ V and $p = 0.324$ mW/μm². The FWHM of the laser spot is 1.44 μm (Supplementary Section I, Fig. S1), which is much smaller than the size of the sample and ensures the spatial resolution. The blue and green arrows denote edges between the sample and electrodes. It is evident that the photocurrent originates from the sample and vanishes at electrodes which excludes the photothermoelectric effect as well as the Schottky barrier origin.

The optoelectronic property of the photoconductive device is further examined by switching, time-dependent and polarization-dependent experiments. Fig. 2f exhibits the on-off repeatability test where the photoresponse remains identical after 2,000 cycles. The period of each cycle is about 10 s. We periodically block the laser and continuously measure the photocurrent versus time. The response speed of the device is found beyond the limit of the repeatability test system, so we perform a modulation-frequency-dependent study to accurately extract the photoresponse time $\tau$ by lock-in technique. The normalized photocurrent at different chopping frequencies is plotted in Fig. 2g. The frequency-dependent photoresponse is expected to follow[44] $I_{ph}(\omega)/I_{ph}(0) = 1/\sqrt{1 + (2\pi\omega\tau)^2}$. The best fitting of the experimental results gives a fast photoresponse time of $\tau = 31.1$ μs. The comparative fast photoresponse suggests finite influence from the dopants. Meanwhile, the Ta$_2$NiS$_5$ crystal is known to be anisotropic, and we studied the photoresponse by illuminating the device with linear-polarized light. The angle-dependent photocurrent is shown in Fig. 2h. With the crystal direction verified by angle-resolved Raman spectra, a prominent anisotropic photocurrent is observed with two-fold symmetry which maximizes along the armchair direction.

The photoconductive origin of the photocurrent is evident by the discussed photocurrent measurements with multiple tuning knobs. The photovoltaic and photothermoelectric mechanisms are firstly ruled out because of the linear bias-dependent and spatial origin. The working frequency of our device also precludes the Dyakonov–Shur mechanism which is usually observed in THz regime[31,45]. The bolometric mechanism behaves similarly in bias- and spatial-resolved experiments, but the response speed of the $Ta_2NiS_5$ device is much faster than the general bolometric devices with a typical response time of $1-100$ ms[46,47]. In addition, the absorption rate of the $Ta_2NiS_5$ is found to be independent of the incident light power. Meanwhile, the conductivity of $Ta_2NiS_5$ increases linearly with temperature. These facts, combined with the observation of superlinear power dependence, further validate that the bolometric effect does not contribute to the observed photoresponse. More details are given in Supplementary Section IV.

In addition to the discussed device performance, an unusual phenomenon is found in the power dependence of the photocurrent. Fig. 3a exhibits $I_{ph} - U$ curves under different incident light intensities. The incident laser spot is kept at the center of the sample. The higher incident power is expected to result in larger photocurrent due to the increased photogenerated electron-hole pairs in the $Ta_2NiS_5$. Lower intensity data is not shown because of the overlapping with other low intensity curves. We extract the photocurrent at $U = 1$ V, as exhibited in Fig. 3b. A clear trend of superlinear power dependence is witnessed. As shown in the inset of Fig. 3b, the photoconductivity first declines with the light power and then increases slowly. A drastic rising of $R_\sigma$ is observed at high illumination power, indicating counterintuitive higher optoelectronic efficiency. To better resolve that, we plot the photocurrent in different incident power regimes and perform the power-law fitting in Fig. 3c-e following $I_{ph} \propto p^\gamma$. With incident power density lower than $0.015$ mW/μm$^2$, a sublinear power dependence of the photocurrent is observed with $\gamma = 0.53 \pm 0.03$. The error scale is given by the fitting error. By increasing the illuminating power, the power dependence of the photocurrent experiences a transition from sublinear to superlinear. Within the power regime of $0.015 - 0.2$ mW/μm$^2$, the photocurrent becomes weakly superlinear with $\gamma = 1.15 \pm 0.01$. As light intensity further increases, strong superlinear dependence is found at the high incident power regime with the power exponent of $\gamma = 1.5 \pm 0.1$. A similar trend can also be found in the linear fit of log-log plot (Supplementary Section III). To the best knowledge of the authors, such strong superlinearity is unusual for homogeneous 2D materials. As summarized in Fig. 3f, the superlinear response of homogeneous 2D devices is generally weak with γ value lower than 1.1[11,23,32,48]. Our result of γ = 1.5 represents a giant superlinearity of the photocurrent in $Ta_2NiS_5$ device which enables higher optoelectronic efficiency at high incident power. The *x*-axis is sorted in the order of report time.

To explain the superlinear dependence of the photocurrent under high incident power, we provide a two-recombination-center (RC) model as illustrated in Fig. 4. Different from previous reports[11,23,49–51] where three centers are required, we will discuss later that the two-RC model is more suitable for narrow gap $Ta_2NiS_5$. The VB and CB denote valence band and conduction band, respectively. Besides, there might also exist a few in-gap states. The presence of those in-gap states is also evidenced by our infrared spectroscopy measurements as discussed in detail in Supplementary Section VI. Our density functional calculation (DFT) suggests that one of the in-gap states might originate from the S vacancy. The in-gap states might also result from other defects such as impurities and dangling bonds[49,52–54] (More details in Supplementary Section VII). These in-gap states act as recombination centers of the photogenerated carriers which could significantly reduce the quantum efficiency. Based on our infrared and transport results, the dopants are at least partially ionized (Supplementary Section VIII). To account for the superlinearity, those two recombination centers ($RC_i$, *i*=1,2) feature distinct parameters. Among them, the most critical two are the density of states ($N_i$) and the capture cross-sections for electrons ($S_{n_i}$). $S_{n_i}$ describes the ability of $RC_i$ to capture the electron. Considering the described system at equilibrium, upon absorbing incident photons, electron-hole pairs are generated across the band gap (process A). The solid and hollow dots denote electrons and holes, respectively. Before those carriers are collected by the electrodes, there is a certain probability (mainly determined by $S_{n_i}$) for the photogenerated electrons to be captured by $RC_1$ (process B) or $RC_2$ (process C). Similarly, RC might also capture the photogenerated holes from valence band (process E & G). Meanwhile, it is possible for the captured electron on $RC_2$ to be thermally emitted to the conduction band (process D) before recombined with holes, while the captured holes might experience the similar procedure (process F). All those procedures influence the carrier population of the states and in turn vary the probability of each procedure. The probability between different processes varies dramatically with different orders of magnitude. For example, the cross-sections of process B, E, H (labeled in dashed line) are negligibly low due the large energy difference between initial and final states. All processes are considered in the model calculation despite its probability. It is worth to note that process A is the only one originating from the photoelectric transition. All remaining processes denote pure electric processes including thermal excitation, trapping, and nonradiative recombination. Other photoelectric transitions and occupation conditions are further discussed in Supplementary Section V. Based on the modulation frequency dependent measurement, all those procedures and resultant carrier population reaches equilibrium within hundred μs. Owning to the orders of magnitude higher capture rate for hole, the photocurrent is dominated by carrier concentration of photogenerated electron $n$ in the conduction band[13,24]. Therefore, the photocurrent reads as $I_{\text{ph}} = nq\mu ES$, where $q$ is electronic charge, $\mu$ is mobility, $E$ is electric field, and $S$ is cross-sectional area of the channel[19]. Without light illumination, the Fermi level in our model lies near the center of gap. This is further supported by the excitation energy extracted by the transport measurement (Details given in Supplementary Section VIII). The Fermi level may stay close to the gap center but deviate a few meV. As a result, $RC_1$ is almost filled and $RC_2$ is nearly empty because of finite thermal excitation. By applying incident light, process C and G are significantly enhanced. Therefore, as reaching the equilibrium shown in Fig. 4a, the $RC_2$ becomes more occupied but most of the states remain empty. Meanwhile, $RC_1$ is less occupied. With higher incident power as depicted in Fig. 4b, the photogenerated carriers lead to the higher occupation rate of $RC_2$ and lower occupation for $RC_1$ which now qualitatively varies the system behavior. The occupation condition influences the strength of all the discussed process A-H.

The response of the photoconductive device can be analyzed by the proposed model. In the upper panel of Fig. 5a, we first consider the conventional case where the properties of $RC_1$ and $RC_2$ are similar. Since the photocurrent is determined by electron concentration, we focus on electron-related processes. Both $RC_1$ and $RC_2$ provide efficient recombination channels through process B and process C, resulting in the recombination of photogenerated carriers before being collected by electrodes. The recombination rate increases with incident power, which in general case, saturates the photocurrent. Therefore, the photocurrent is expected to be linear or sublinear on the power dependence, as exhibited in the lower panel. As suggested by previous work[13], different capture cross-sections of the RC potentially lead to weak superlinear dependence of the photocurrent. If the electron capture cross-section of RC₁ ($S_{n_1}$) is much smaller than RC₂ ($S_{n_2}$), the recombination channel on the RC₁ is effectively closed (denoted by red crossings in Fig. 5b) and most of the photogenerated electron is trapped by RC₂. At high incident power regime, the RC₂ becomes densely occupied, which suppresses the process C as denoted by dashed line. Lower recombination rate at higher intensity allows for higher optoelectronic efficiency and leads to the superlinear behavior of the power dependence. To account for the observed giant superlinear photoresponse, another critical parameter $N_i$ is taken into consideration. With much lower density of states of RC₂ shown in Fig. 5c, the occupation condition qualitatively varies at high incident power. Since the density of states of RC₂ is much lower, higher incident light leads to the rapid saturation of RC₂ which forbids the C process as well as the recombination channel on RC₂. Combined with negligible $S_{n_1}$, now both of the recombination channels are closed at the high power regime. Therefore, a giant superlinear power dependence is presented as shown in the lower panel.

To elucidate the effects of $N_1/N_2$ on superlinear photocurrent, we perform the numerical calculation based on the two-RC model. For each energy level in this model, all related carrier procedures reach equilibrium in the end. For example, the photogenerated electron concentration of the conduction band is given by

$$\frac{dn}{dt} = F - n[vS_{n_1}(N_1 - n_1) + vS_{n_2}(N_2 - n_2)] + n_2 P_2 - S'vnp = 0 \quad (4)$$

The $F$, $-nvS_{n_1}(N_1 - n_1)$, $-nvS_{n_2}(N_2 - n_2)$, $n_2 P_2$, and $-S'v'np$ term corresponds to the procedure A, B, C, D, and H, respectively; $n, n_1, n_2$ represents the electron density of conduction band, RC₁, and RC₂, respectively; $p$ represents the hole density of valence band; $v$ denotes thermal velocity of the carriers which is assumed to be equal for simplicity; $F$ denotes the density of electron-hole pairs created by optical excitation per second which is determined by light intensity, quantum efficiency, and absorption rate; $S_{n_1}$, $S_{n_2}$ denotes the electron capture cross-section of RC₁ and RC₂, respectively; $S'$ denotes recombination cross-section between free electrons and free holes; $P_2$ denotes the probability per unit time for the thermal ejection of an electron in RC₂ into the CB. For all other energy levels, similar equations can be derived by fully considering related carrier procedures which ultimately achieve equilibrium. The overall equations are provided in Supplementary Section V. By solving the nonlinear equations, the photoresponse of the device with different parameter settings can be numerically extracted. Fig. 5d depicts the occupation proportion of RC$_2$. For parameter settings with high $N_1/N_2$ (green curve), the electron concentration is intensely saturated at high incident power. In contrast, such a saturation feature is weakened for lower values setting of $N_1/N_2$. The difference in the electron density has a profound influence on the electron density of conduction band through process C and process D. This is further supported by the calculated photocurrent in Fig. 5e. Regardless of the $N_1/N_2$ setting, all curves exhibit a superlinear feature due to the negligible $S_{n_1}$. However, a higher $N_1/N_2$ leads to a more prominent superlinear power dependence of the photocurrent which agrees with the discussed picture.

To quantitively verify the proposed model, we perform the multi-parameter fitting as shown in Fig. 5f. Due to the nonlinearity of the equation, the fitting is carried out using gradient descent method which reaches convergence within 1 day. The experimental available values, such as the gap size, are fixed. The black dots denote experimental data of the Ta$_2$NiS$_5$ device which is well-fitted by the model (red line). The critical fitting parameters are extracted as $N_1/N_2 = 16.06$, $S_{n_1}/S_{n_2} = 0.87 \times 10^{-4}$ (more details are provided in Supplementary Section V). It is worth noting that the experimental data can be fitted by both two-RC and three-RC models with qualitative similarity. Thus, the two-RC model is introduced for simplicity which also avoids possible overfitting. Here the fitted model describes a small gap semiconductor system in agreement with our infrared spectroscopy result and the generally accepted picture[35,36,55]. Further wavelength-dependent research, especially in the mid-infrared regime, might give new insights into both band information and photoresponse.

The observed fast photoresponse speed also agrees with the proposed model. The superlinear photocurrent requires the presence of recombination centers. However, the strong superlinearity also requires the dopant density as well as the density of states for the in-gap states to be low. Only with low dopant density, the recombination center can be fully occupied at high illuminating power. Otherwise, superlinearity is not expected to be observed. The fast photoresponse speed is further contributed by the suppression of the recombination process under light illumination. While lowering the power, the response speed of the device drops as discussed in detail in Supplementary Section IX. As a result of superlinearity, the photocurrent is also expected to drop while expanding the laser spot. The deduction has been confirmed in our beam-size-dependent experiments as discussed in detail of Supplementary Section X.

To test the Ta$_2$NiS$_5$ photodetector and observed superlinearity for potential applications, the imaging function of the device is evaluated. The fabricated device is transferred to the image plane of a camera and controlled by a translation stage so that it mimics the CCD of the camera after a complete scanning. A screen showing the image of an apple is used as the target with tunable brightness. The photos are exhibited in Fig. 6a with 100 ms exposure time where the apple is successfully photographed. With higher brightness of the target, the apple becomes more distinguishable and can be observed in more detail. For more quantitative analysis, root mean square (RMS) contrast of the image is extracted with the definition

$$\text{RMS contrast} = \sqrt{\frac{1}{MN} \sum_{i=0}^{M-1} \sum_{j=0}^{N-1} (I_{\text{ph}ij} - \bar{I}_{\text{ph}})^2}, \tag{5}$$

where $M$ and $N$ are the number of pixels per row and per column, respectively. $\bar{I}_{ph}$ is the average value of the signal. As shown in Figure. 6b, the RMS contrasts of the image increase with the maximum detected power among all pixels. Notably, a superlinear trend is found, resembling superlinear photoconductivity. By supporting better imaging contrast, superlinear photodetector could be promising for future optoelectronic detection.

The recombination center plays an important role in the superlinearity photoresponse. More sophisticated experimental tools and theoretical calculations might help to identify the origin of the in-gap states and extract their evolution upon light illumination for better understanding of the system.

In summary, we report the optoelectronic characteristics of the photoconductive detector based on multi-layer Ta$_2$NiS$_5$ and discover a giant superlinear power dependence of photocurrent. The time-resolved, frequency-resolved, spatial-resolved, bias-resolved, and angle-resolved photocurrent measurements not only present a fast, endurable, and anisotropic photoresponse, but also suggest the photoconductive nature of the device which ensures that the device performance is determined by the material property of $Ta_2NiS_5$. Starting from illumination power density of 1.54 μW/μm², photocurrent presents a sublinear photocurrent with the power density. Around 15.4 μW/μm², a transition from sublinearity to superlinearity is witnessed. With incident power density higher than 0.2 mW/μm², a prominent superlinearity is observed with power exponent γ = 1.5. The strong superlinearity can be quantitatively explained by a two-RC model. The in-gap recombination centers with distinct density of states and capture cross-sections lead to the rapid saturation of carrier occupancy, thereby, both recombination channels are closed and enable higher optoelectronic efficiency at large incident power. The quantitative fitting between the proposed model and experiments further validates the proposed physical mechanism. The photos taken by the $Ta_2NiS_5$ demonstrate enhanced RMS contrast showing potential applications of the superlinear photocurrent. Our work paves the way for the superlinearity of optoelectronic devices and enables better device performance in high-power applications.

## Methods

### Crystal growth and characteristic

$Ta_2NiS_5$ single crystals were prepared by standard chemical vapor transport method. Stoichiometric mixture of Ta, Ni, and S powder was sealed in a 20 cm vacuum quartz tube with iodine as transport agent. The tube was loaded in a two-zone furnace kept at 950 ℃ and 830 ℃. After a 5-day growth procedure, shiny and needle-like single crystals were found in the low-temperature zone. X-ray diffraction was tested by Bruker D8 Discover. Raman spectrum was tested by a home-built system using 632.8 nm laser.

### Device fabrication

Multi-layer $Ta_2NiS_5$ flakes were mechanically exfoliated from bulk crystals and then transferred to a $Si/SiO_2$ wafer. Devices were fabricated by self-made lithography system with lift-off procedures. Cr/Au (5 nm/70 nm) is deposited as electrodes and Ohmic contact has been proved by the $I - U$ measurement.

### Photocurrent measurement

Photodetectors were excited by 632.8 nm laser through a 50×, NA = 0.8 objective. The FWHM of the focal spot is 1.44 μm. Bias-, angle-, spatial-, time- and power-dependent photocurrent were measured by Keithley 2450 using two-terminal method. Spatial photocurrent scanning was carried out by an additional piezo-actuated stage. The photoswitching test was performed with a periodically switched shutter. The modulation frequency-dependent measurement was carried out by Stanford Research SR-860 lock-in amplifier with a chopper.

**Supporting Information**

Supplementary information is available from the Wiley Online Library or from the author.

**Acknowledgements**

We gratefully thank T.-Y. Zhai, L.-J. Li, Q.-J. Wang, J.-B. Li, Y.-Z. Zhang, and L. Li for helpful suggestions on the superlinearity photocurrent and experiments. X.Y. was supported by the National Natural Science Foundation of China (grant no. 62005079 and no. 12174104), the Shanghai Sailing Program (grant no. 20YF1411700), the International Scientific and Technological Cooperation Project of Shanghai (grant no. 20520710900) and a start-up grant from East China Normal University. C.Z. was supported by the National Natural Science Foundation of China (grant no. 12174069), Shanghai Sailing Program (grant no. 20YF1402300), Natural Science Foundation of Shanghai (grant no. 20ZR1407500), the Young Scientist project of the Ministry of Education innovation platform and a start-up grant from Fudan University. NBJ was supported by the Prime Ministers Research Fellowship. A.N. thanks the Indian Institute of Science for a startup grant. C.D. and F.Y. acknowledge the support from the National Key Research Project (grant 2022YFA1402902) and the National Natural Science Foundation of China (grant 62274061).

**Conflict of Interest**

The authors declare no conflict of interest.

**Author contributions**

X.Y. conceived the idea and supervised the overall research. X.M. carried out the growth of the $Ta_2NiS_5$ single crystals; J.M. and Y.D. performed the crystal characterization including XRD and infrared spectrum with help from W.W. and Z.S. under the supervision of C.Z., C.D. and Z.S.; X.D., J.W., B.L., and Y.M. conducted the device fabrication under the supervision of F.Y., N.Z., P.X., and C.D.; X.M. conducted the photocurrent experiments with help from Y.D. and Z.S.; N.J. performed the DFT calculation under the supervision of A.N.; X.Y., X.M., Y.D., C.Z., and J.C. wrote the paper with the help of all authors.

**Data availability statement**

The data that support the findings of this study are available from the corresponding author upon reasonable request.

**Keywords**

layered ternary chalcogenides, superlinear photoresponse, photoconductive detector, high power sensor

# Figures

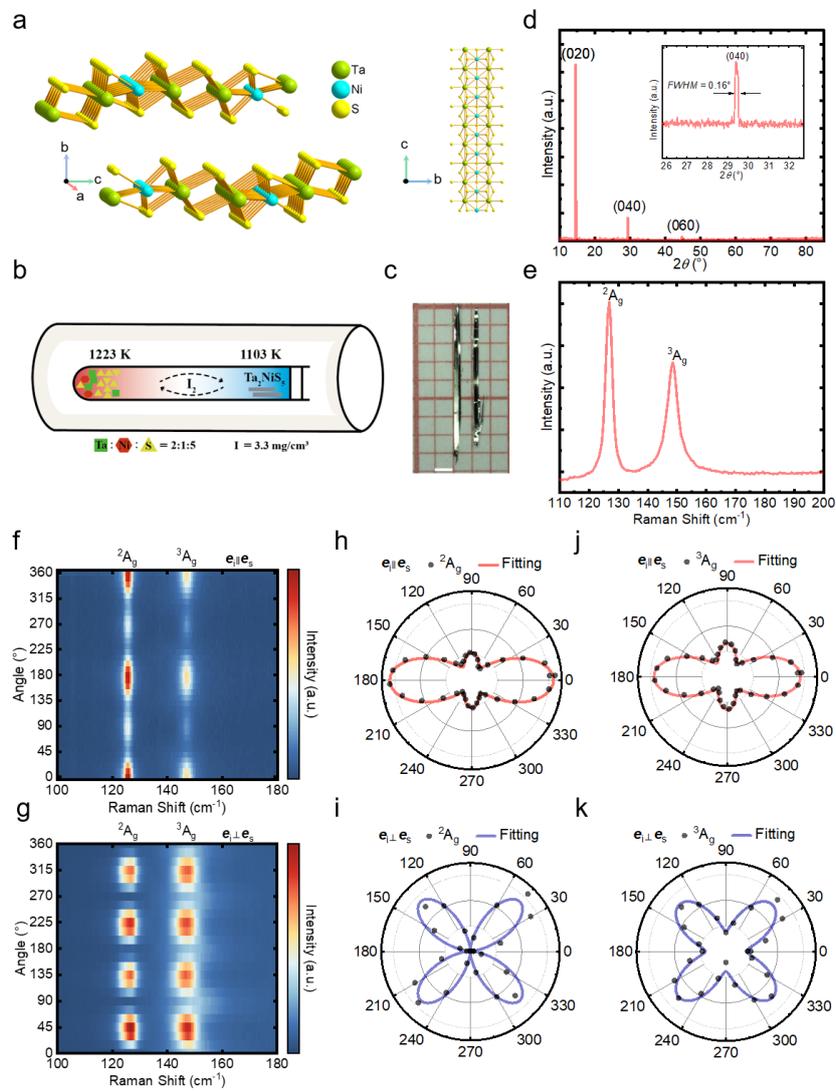

**Fig. 1 | X-ray and Raman spectrum of Ta₂NiS₅ single crystals. a**, The crystal structure of Ta₂NiS₅. **b**, Schematic diagram of the chemical vapor transport process. **c**, The photo of the as-grown single crystal. The scale bar is 1 mm. **d**, X-ray diffraction pattern of Ta₂NiS₅. **e**, Raman spectrum of Ta₂NiS₅ with random polarization. **f-k,** False-color maps and corresponding intensity fittings of the polarization-dependent Raman spectra in parallel and perpendicular configurations.

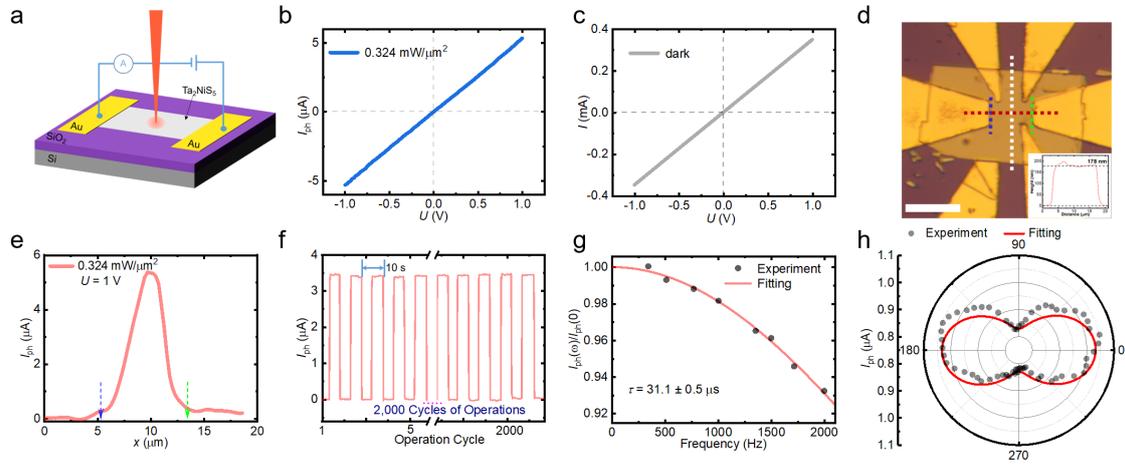

**Fig. 2 | Photoconductive origin and performance of the photodetector. a**, Schematic diagram of the optoelectronic device structure based on exfoliated Ta$_2$NiS$_5$. **b**, Linear bias dependence of the photocurrent. **c**, Linear dark current which suggests Ohmic contact of the device. **d**, The photo of the device. The scale bar is 10 μm. The inset exhibits the height profile of the sample (measured along the white dashed line). **e,** Photocurrent profile taken along the red dashed line in **d**. The blue and green arrows denote the edges between the sample and the electrode. **f**, Photoswitching test. **g**, Frequency-dependent photocurrent measurement of the Ta$_2$NiS$_5$ device. The fitted response time is $\tau = 31.1 \pm 0.5$ μs. **h**, Polarization-dependent photocurrent of the Ta$_2$NiS$_5$ device.

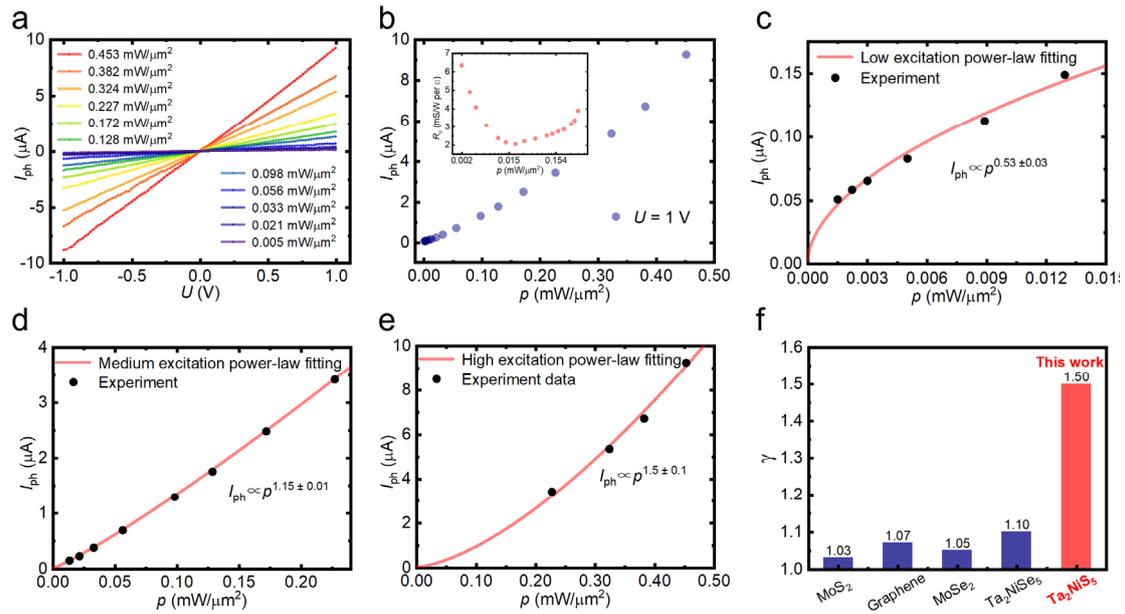

**Fig. 3 | Superlinear power dependence of photocurrent. a**, $I_{ph} - U$ curves of the device under different illuminating power. **b**, The photocurrent extracted at 1 V bias. The inset is the $R_\sigma$ with a counterintuitive V-shape. Most of the reported optoelectronic devices exhibited a monotonical decrease. **c-e,** The photocurrent and power law fitting at different incident power regimes. The power dependence varies from sublinear at low intensity to superlinear at high intensity. The $\gamma = 1.5$ is reached after 0.2 mW/μm². **f,** A comparison of superlinear power exponent among various homogeneous 2D devices.

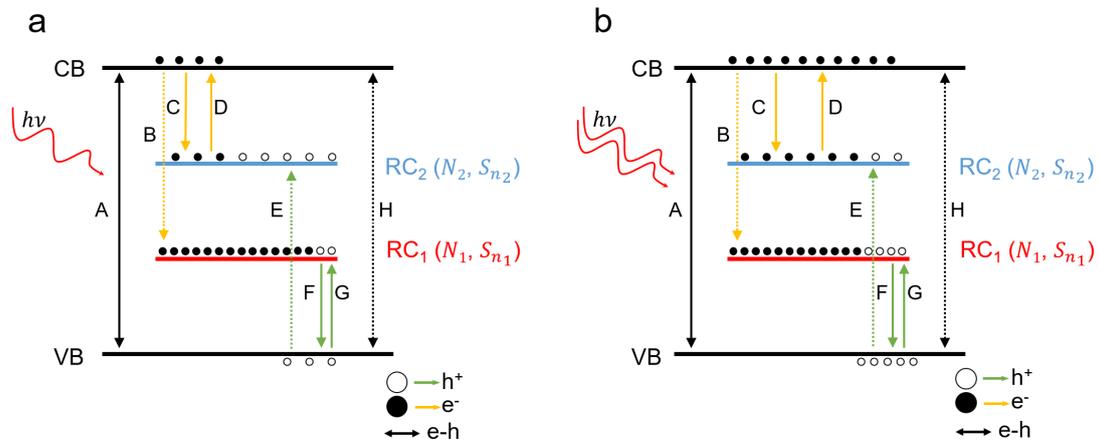

**Fig. 4 | Schematic plot of the two-RC model. a,** Band diagram at low illumination power. Two recombination centers (RC) are characterized by different density of states ($N_i$) and electron capture cross-sections ($S_{n_i}$). The black arrows refer to electron-hole pairs generation (A) due to the photoexcitation and recombination between free electrons and holes (H). A series of additional carrier processes coexist for both electrons (yellow arrows) and holes (green arrows), including trapping, thermal excitation, and recombination. Solid and hollow dots denote the distribution of electrons and holes on each band, respectively. At low illumination power, the $RC_2$ is partially filled by electrons which favors the capture and recombination processes of the photogenerated carriers. Note that the schematic plot of the energy level is not to scale. Process A is the only process from photoelectric transition while all others are electric processes. The dashed lines indicate the transitions with low probability. **b,** Band diagram at high illumination power. The major states of $RC_2$ become occupied which suppresses the electron trapping process. Thus, the recombination rate drops and potentially leads to superlinear power dependence of the photocurrent.

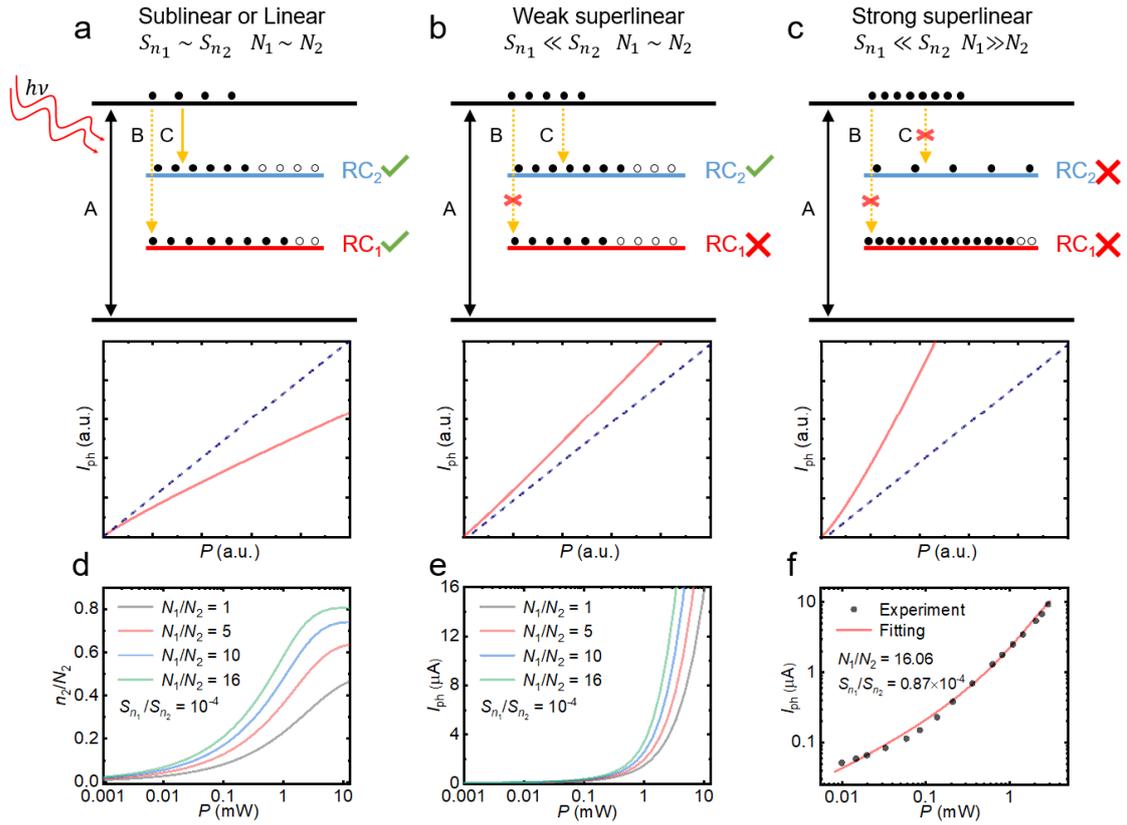

**Fig. 5 | Two-RC model with different parameters and fitting to the experimental data. a-c,** The upper panels denote the carrier distribution on the recombination centers at high illumination power. The lower panels denote the corresponding power dependence of the photocurrent. In the left panel ($S_{n1} \sim S_{n2}$, $N_1 \sim N_2$), both in-gap states work as efficient recombination centers, leading to the sublinear or linear photoresponse. In the middle panel ($S_{n_1} \ll S_{n_2}$, $N_1 \sim N_2$), the negligible electron capture cross-section of the lower in-gap state closes the recombination channel on the RC$_1$. Combined with the slow saturation of RC$_2$ at the high power regime, weak superlinear power dependence is achieved. In the right panel ($S_{n_1} \ll S_{n_2}$, $N_1 \gg N_2$), the lower density of states of upper in-gap state results in a rapid saturation which effectively closes both recombination channels and potentially leads to prominent superlinear photoresponse. **d,** The calculated occupancy ratio of RC$_2$ based on the two-RC model. A higher ratio of $N_1/N_2$ leads to a rapid saturation. **e,** The calculated power dependence of the photocurrent. The photoresponse features more prominent superlinearity with a higher $N_1/N_2$ ratio. **f,** The fitting to the experimental data. The two-RC model fits well with the experimental result.

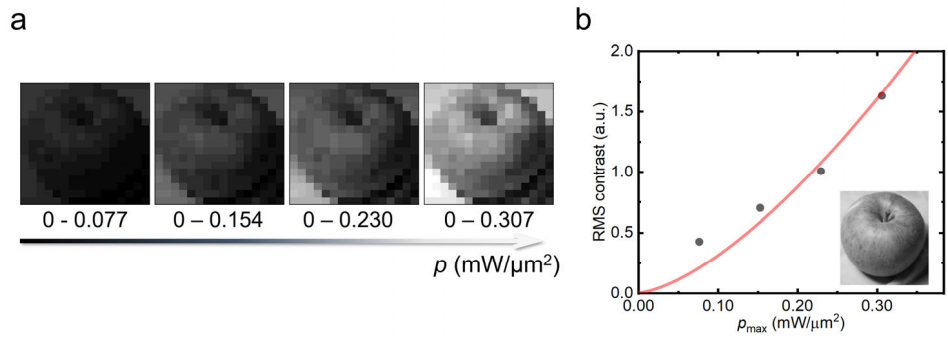

**Fig. 6 | Photography by the Ta$_2$NiS$_5$ device and the image contrast. a,** The images of an apple are successfully photographed which are more distinguishable with higher brightness of the target. **b,** The RMS contrast is enhanced with power following a superlinear trend. The inset is the original target.